\documentclass[a4paper,11pt]{article}
\pdfoutput=1 

\usepackage{jcappub} 

\usepackage[T1]{fontenc} 
\usepackage{enumitem}
\usepackage{xcolor}

\title{Unexplained correlation between the Cosmic Microwave Background temperature and the local matter density distribution}

\author[a,1]{M. Cruz\note{Corresponding author.}}
\author[a]{E. Martínez-González,}
\author[a,b]{C. Gimeno-Amo,}
\author[a]{B. J. Kavanagh}
\author[c]{and M. Tucci}

\affiliation[a]{Instituto de Física de Cantabria (CSIC-Universidad de Cantabria),\\Avda. de los Castros s/n, 39005 Santander, Spain}
\affiliation[b]{Departamento de Física Moderna, Universidad de Cantabria,\\Avda. de los Castros s/n, 39005 Santander, Spain}
\affiliation[c]{Department of Astronomy, University of Geneva,\\ch d'Écogia 16, 1290 Versoix, Switzerland}

\emailAdd{cruz@ifca.unican.es}
\emailAdd{martinez@ifca.unican.es}
\emailAdd{gimenoc@ifca.unican.es}
\emailAdd{kavanagh@ifca.unican.es}
\emailAdd{marco.tucci@unige.ch}

\abstract{
Recent observations have indicated a Cosmic Microwave Background (CMB) temperature decrement in the direction of local galaxies within the 2MASS Redshift Survey. We investigate this detection by analyzing its frequency dependence and sensitivity to component separation methods, suggesting that Galactic foregrounds are unlikely to be the cause. Contrary to previous studies, we find that the decrement is independent of galaxy type, indicating a possible correlation between the CMB and the overall matter density field. To test this hypothesis, we employ three analytical approaches: cross-correlation analysis, template fitting, and Bayes Factor calculation. Our cross-correlation analysis shows a significant correlation ($p < 0.7\%$) between the CMB and the 2MASS Redshift Survey projected matter density at distances below 50 Mpc/$h$. Template fitting and Bayes Factor analyses support this finding, albeit with lower significance levels ($1\% - 5\%$). Importantly, we do not detect this signal beyond 50 Mpc/$h$, which constrains potential physical interpretations. We discuss that the physical origin of this correlation could potentially be linked to the dark matter distribution in the halos of galaxies. Further investigation is required to confirm and understand this intriguing connection between the CMB and local matter distribution.}

\begin{document}
\maketitle
\flushbottom

\section{Introduction}
\label{sec:intro}

The Cosmic Microwave Background (CMB) is a cornerstone of modern precision cosmology~\cite{Turner:2022gvw}. The CMB has an almost perfect black-body spectrum with a temperature of $T_\mathrm{CMB} = 2.725\,\mathrm{K}$~\cite{2009ApJ...707..916F}, upon which tiny fluctuations at the level of $\Delta T/T \sim 10^{-5}$ are imprinted. These anisotropies provide a detailed snapshot of the Universe approximately 380,000 years after the Big Bang, but also provide a wealth of information about the intervening Universe, crossed by the photons from the surface of last scattering to today. Combined with other cosmological probes, the analysis of the CMB anisotropies has allowed us to infer the geometry, dynamics and composition of the Universe~\cite{Planck:2018nkj,Planck:2018vyg}.

The CMB anisotropies can be broadly classified as either primary or secondary. Primary anisotropies arise from the surface of last scattering and reflect the conditions of the early Universe. In contrast, secondary anisotropies arise from interactions between CMB photons and large-scale structures on their subsequent journey to us. 
Among the sources of secondary anisotropies, notable examples include the Integrated Sachs-Wolfe (ISW) effect \cite{1967ApJ...147...73S}, 
the Sunyaev-Zel'dovich (SZ) effect \cite{1972CoASP...4..173S,1980ARA&A..18..537S,1980MNRAS.190..413S}, and gravitational lensing~\cite{1987A&A...184....1B}. The ISW effect results from the interaction of CMB photons with time-varying gravitational potentials, while the SZ effect is caused 
by inverse Compton scattering of CMB photons by hot electrons in galaxy clusters. Gravitational lensing distorts the CMB pattern due to deflections by intervening mass distributions.

Though the CMB has allowed for sub-percent level inference of cosmological parameters, it has also given rise to a number of unexplained anomalies (see Refs.~\cite{Schwarz:2015cma,Abdalla:2022yfr,Aluri:2022hzs,Planck:2013lks} for reviews). Some of these anomalies appear in the large-scale properties of the CMB, such as the surprising alignment between the CMB quadrupole and octopole~\cite{deOliveira-Costa2004,Copi:2013jna}; the hemispherical power asymmetry~\cite{Eriksen2004,Hansen2004}; the anomalously low variance~\cite{Monteserin2008,Cruz2011}; the lack of power at large scales~\cite{Bernui2006}; and the Cold Spot~\cite{Vielva2004,Cruz2005}. Others arise as a tension with other cosmological constraints, such as the discrepancy between the value of the Hubble constant as inferred from the CMB and from late Universe probes~\cite{Hu:2023jqc}. Exploring these anomalies has the possibility to deepen our understanding of the CMB as well as to point towards possible sources of New Physics. 

Recently, an intriguing CMB temperature decrement has been observed in the direction of local galaxies within the 2MASS Redshift Survey, as reported by~\cite{Luparello2023}. That is, the CMB temperature appears to be lower in the direction of local galaxies.
This phenomenon, which remains unexplained by any known secondary anisotropies, was identified through a stacked profile analysis of selected 2MASS galaxies. The study focused on spiral galaxies with large physical sizes and radial velocities up to $4500$ km/s, revealing a CMB decrement of about $10 \mu$K extending over more than $10$ degrees from the galaxy centres. 

Reference~\cite{Hansen2023} further created a decrement template from the 2MASS Redshift Survey data, 
suggesting that this signal could explain some of the CMB anomalies, such as the Cold Spot ~\cite{Lambas2024}
and the hemispherical power asymmetry. 
However, Ref.~\cite{Addison2024} revisited this analysis, 
arguing that the initial studies significantly underestimated the uncertainties. Consequently, 
they contend that the observed signal is not statistically significant.

Despite these conflicting results, the cosmological implications of the decrement, if confirmed, are profound. A genuine correlation between the CMB and local matter distribution could necessitate re-evaluating cosmological parameters derived from CMB measurements.
Consequently, further rigorous investigation is essential not only to resolve the current debate but also to explore its potential connections to observed CMB anomalies and its broader implications for cosmology.
  
In this paper, we reproduce and extend the analysis conducted by Ref.~\cite{Luparello2023} to examine the 
reported CMB decrement. We explore its frequency dependence, the dependence on the type and size of the galaxies, the influence of different foreground subtraction methods~\cite{PlanckComponentSepatation2020}, and we reevaluate its statistical significance. We also utilize Constrained Realizations of the 2MASS density field provided by \cite{Lilow2021} to generate a projected matter density field, and perform a cross-correlation analysis between the CMB and the projected matter density field.
Building on this analysis, we treat the projected matter density field as a template and conduct a template fitting analysis to determine its best-fit amplitude and corresponding Bayes factor. 
This comprehensive approach aims to shed light on the nature of the CMB decrement and its possible implications for our understanding of the local Universe and CMB anomalies.

This paper is structured as follows: Section~\ref{sec:data} describes the \emph{Planck} data and 2MASS Redshift Survey data utilized in our study, and outlines our method for generating the projected density field of galaxies from the 2MASS Redshift Survey data. Section~\ref{sec:radial_profiles} details our reproduction of the analysis by~\cite{Luparello2023}, examining the frequency dependence, galaxy size and type dependence, and the impact of different foreground subtraction methods. In this section, we also analyze the statistical significance of the observed temperature decrement, providing a quantitative assessment of the robustness of the signal. 
Section~\ref{sec:crosscorrelation} presents our cross-correlation analysis between the CMB and the projected matter density field. Section~\ref{sec:TemplateFit} details our template fitting analysis, determining the best-fit amplitude of the template and evaluating the Bayesian evidence for its presence, comparing these findings with corresponding simulations. Section~\ref{sec:Discussion} explores possible physical origins of the observed effect. Finally, Section~\ref{sec:conclusions} summarizes our conclusions. 

\section{CMB, galaxy and density maps}
\label{sec:data}
Our analysis primarily employs the 2018 \emph{Planck} collaboration CMB data products, known as PR3~\cite{Planck2018}. This dataset includes maps cleaned using four distinct component separation methods~\cite{PlanckComponentSepatation2020}: Commander~\cite{Commander2008}, NILC (Needlet Internal Linear Combination)~\cite{NILC2012}, SEVEM~\cite{SEVEM2012}, and SMICA (Spectral Matching Independent Component Analysis)~\cite{SMICA2008}. The common mask ~\cite{PlanckComponentSepatation2020} is applied across all maps to exclude the Galactic plane and regions with potentially high foreground contamination. PR3 also provides 300 simulated CMB maps with corresponding End-to-End noise simulations for each component separation method.
For frequency-specific analysis, we use SEVEM frequency maps from the PR4 data release~\cite{Planck2020}, covering the 70, 100, 143, and 217 GHz channels. PR4 does not provide maps for all component separation methods, but represents the most recent dataset which contains SEVEM clean maps at different frequencies. Additionally, we incorporate a 44 GHz SEVEM clean map from the DX9 \emph{Planck} intermediate results~\cite{Barreiro}.

The 2MASS Redshift Survey (2MRS)~\citep{Huchra2012} is currently the most complete all-sky redshift survey of the local universe. It contains $44599$ galaxies with redshifts, $z$, up to $0.1$, providing comprehensive data including isophotal radii and morphological types for a virtually complete sub-sample of 20,860 objects with $K \leq 11.254$ mag and $|b| > 10$ degrees. 
Galaxy types are encoded using numeric codes representing various classifications, as detailed in Section 5 of~\cite{Huchra2012}.
For our analysis, we group the galaxy types into broader categories:
\begin{itemize}
  \item elliptical galaxies  (codes -7 to -5),
  \item lenticular galaxies (-4 to -1),
  \item early-type spirals (0 to 2),
  \item late-type spirals (3 to 7),
  \item other types, including spirals with most pronounced disks, irregular, peculiar and unclassified galaxies (8 to 98).
\end{itemize}
This grouping facilitates the statistical analysis, ensuring that each group has more than $700$ galaxies for the considered redshift range.

Ref.~\cite{Lilow2021} provide a constrained realization of the 2MRS matter density and peculiar velocity fields, 
generated using the COnstrained Realizations from All-sky Surveys (CORAS) methodology. CORAS is based on applying a Wiener Filter to a spherical Fourier-Bessel expansion of the density field, suitable for a flux-limited all-sky redshift survey. It assumes the matter density contrast and the comoving peculiar velocity to be related via linear theory.

The reconstructed normalized density contrast is discretized on a regular $201^3$ Cartesian grid in comoving Galactic coordinates, with each coordinate running from $-200$ Mpc/$h$ to $+200$ Mpc/$h$ in steps of $2$ Mpc/$h$. 
The normalized density contrast is the density contrast divided by $\sigma_8$. Note that the uncertainties in the normalized density contrast show a radial increase from the origin towards the boundary of the reconstruction volume \cite{Lilow2021}.
While the CORAS data will play a key role in our correlation analysis in Section \ref{sec:crosscorrelation}, we first proceed with the galaxy-based analysis in the following section.

\section{Mean temperature radial profiles around nearby galaxies}
\label{sec:radial_profiles}

The CMB decrement reported by~\cite{Luparello2023} was detected through a stacking analysis of CMB temperature maps centered on nearby galaxies. This method involves calculating the mean radial temperature profile, $\langle \Delta T(\theta) \rangle$, as a function of the angular distance $\theta$ from the galaxy centers. We define this profile as:
\begin{equation}
\langle \Delta T(\theta) \rangle = \frac{\sum_{k=1}^{N_{\mathrm{gxs}}} \sum_{i \in C_k(\theta)} \Delta T_{k, i}(\theta)}{\sum_{k=1}^{N_{\mathrm{gxs}}} N_k(\theta)},
\label{eq:RadialProfile}
\end{equation}
where $i$ and $k$ index the CMB map pixels and galaxies, respectively. The inner sum is performed over the set $C_k(\theta)$, comprising all $N_k(\theta)$ pixels within the annulus $(\theta, \theta + \Delta\theta)$ centered on each of the $N_{\mathrm{gxs}}$ galaxies in the sample. 
This formulation adapts Eq.~(1) from~\cite{Luparello2023} to account for potential variations in $N_k(\theta)$ due to regions excluded by the common mask. By weighting the mean by $N_k(\theta)$, we ensure a more accurate representation of the average temperature decrement, particularly in partially masked regions.
\begin{figure}[tb]
  \centering
  \includegraphics[width=0.8\textwidth]{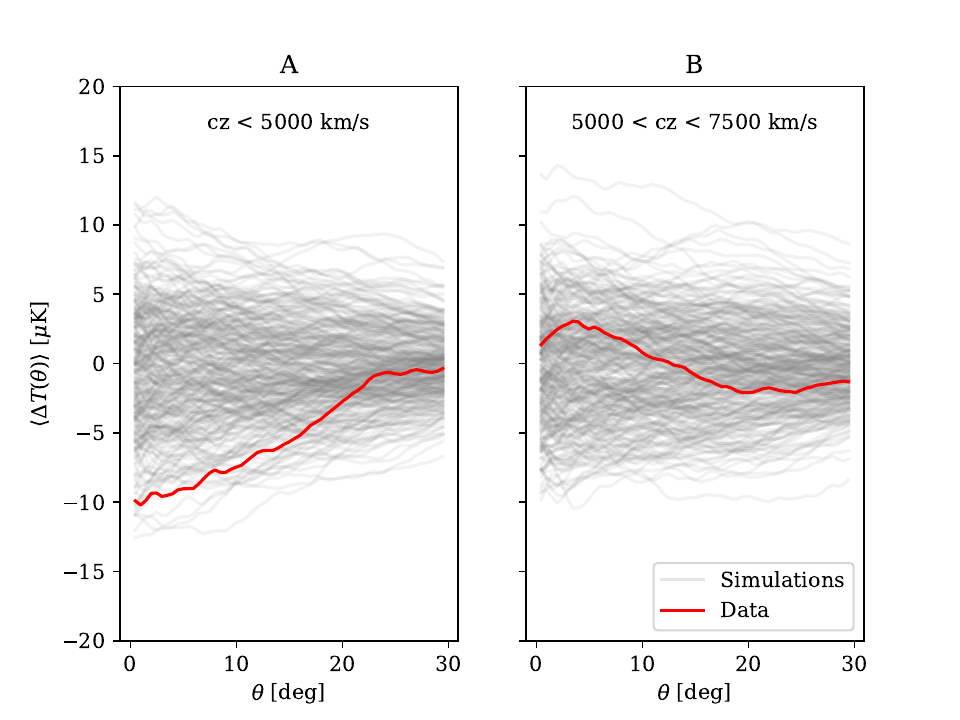}
  \caption{Mean radial profiles for SMICA PR3 data (red) and $100$ simulations (grey) as a function of angular distance, $\theta$, from the galaxy centers. Galaxies with radial velocities in the range $0 < cz < 5000$ km/s (panel A) and $5000 < cz < 7500$ km/s (panel B) are considered. \label{fig:radial_profiles_sims}}
\end{figure}
Previous studies have employed varying approaches to scale selection in their analyses. References~\cite{Luparello2023} and~\cite{Addison2024} considered scales down to 0.01 degrees, while~\cite{Hansen2023} used $\theta \geq 0.2$  degrees, since the \emph{Planck} point source mask removes a disc of radius $0.2$ degrees around point sources due to possible contamination. In our analysis, we adopt the more conservative criterion of~\cite{Hansen2023}, setting $\theta_{\mathrm{min}} = 0.2$ degrees with a ring width of $\Delta \theta = 0.5$ degrees. This choice allows us to use a HEALPix\footnote{http://healpix.sourceforge.net} resolution parameter of $N_{\mathrm{side}} = 256$, which is sufficient for our mean radial profile analysis while minimizing computational demands.
Regarding the redshift range,~\cite{Luparello2023} used $cz \leq 4500$, while~\cite{Hansen2023} extended this to $cz \leq 5100$ km/s. Our tests revealed that small variations in this range do not significantly affect the results. Consequently, we opt for an intermediate value of $cz \leq 5000$ km/s for our analysis.

Figure~\ref{fig:radial_profiles_sims} illustrates the mean radial profile for galaxies in the 2MASS Redshift Survey, divided into two subsets: those with radial velocities up to $5000$ km/s (panel A) and those with velocities between $5000$ and $7500$ km/s (panel B). The upper limit of $7500$ km/s was chosen to ensure that both subsets contain a similar number of galaxies, allowing for a balanced comparison. The figure compares observational data with $300$ simulations for SMICA PR3.

Our analysis confirms the $10$ $\mu$K  decrement reported by~\cite{Luparello2023} in the observational data with $0 < cz < 5000$ km/s. However, when viewed in the context of the simulations, the significance of this decrement appears less definitive. Despite the absence of intrinsic correlations between the simulated CMB maps and the observed galaxy distribution, some simulations exhibit chance correlations that produce signals comparable to those observed in the data. This underscores the importance of rigorous statistical analysis to distinguish genuine signals from chance alignments.
Moreover, we confirm that this signal is exclusively present for galaxies with radial velocities in the range $0 < cz < 5000$ km/s. We examined various other radial velocity ranges beyond $5000$ km/s (e.g., $5000$ to $7500$ km/s, $7500$ to $10000$ km/s, $5000$ to $10000$ km/s, $10000$ to $20000$ km/s, among others) and found no significant decrement in any of these higher velocity bins. 

To explore whether this observed decrement could be caused by Galactic foregrounds, we examine its consistency across different component separation methods and frequency bands. 
\begin{figure}[tb]
  \centering
  \includegraphics[width=0.8\textwidth]{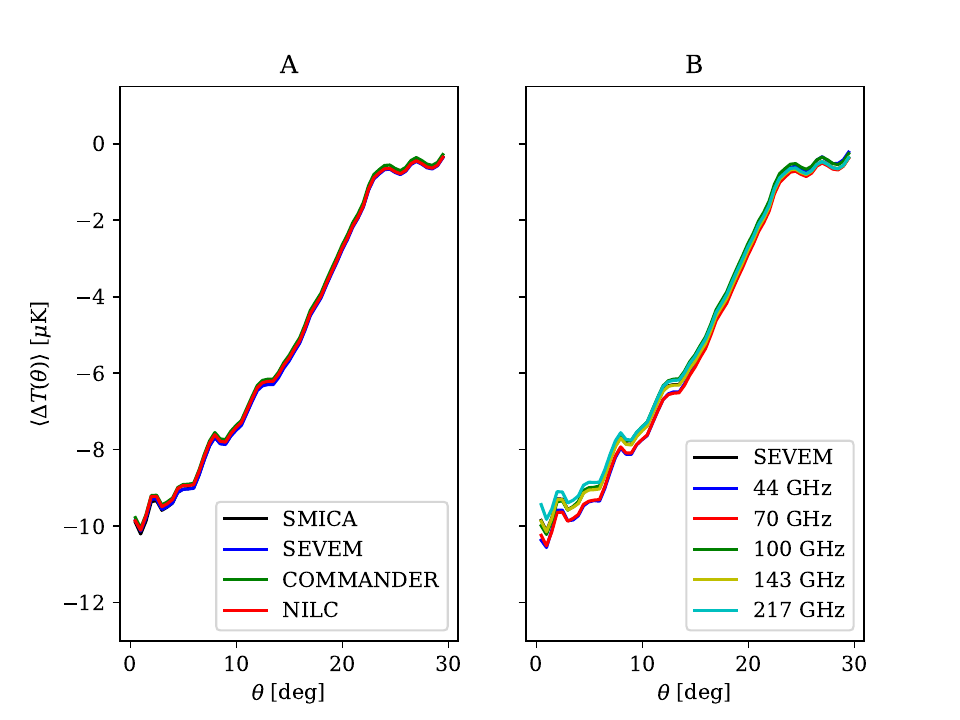}
  \caption{Panel A: Mean radial profiles for different PR3 component separation methods. Panel B: Mean radial profiles for SEVEM frequencies 44, 70, 100, 143, 217 GHz.
  Galaxies with radial velocities up to $5000$ km/s are considered in both panels.
  \label{fig:radial_profiles_methods_freq}}
\end{figure}
Figure~\ref{fig:radial_profiles_methods_freq}A shows a comparison of mean radial profiles across different component separation methods. Our analysis reveals that the CMB decrement signal is consistently present in all methods, with remarkably similar radial profiles. This strong consistency across SMICA, NILC, SEVEM, and Commander methods is particularly noteworthy, as it contrasts with the findings of~\cite{Addison2024}, who reported significant differences between SMICA and SEVEM results.
The uniformity in our results across different component separation techniques suggests that the observed decrement is unlikely to be an artifact of a specific foreground removal method.

Moreover, Figure~\ref{fig:radial_profiles_methods_freq}B demonstrates remarkably consistent radial profiles across SEVEM frequency maps at 44, 70, 100, 143, and 217 GHz. This frequency independence of the signal is a crucial finding, as it strongly suggests that Galactic foregrounds are not responsible for the observed CMB decrement. Galactic contaminants typically exhibit distinct spectral signatures, which would manifest as variations in the signal strength across different frequencies. The absence of such variations is then strongly suggestive of a cosmological origin for the CMB decrement.

\begin{figure}[tb]
  \centering
  \includegraphics[width=0.8\textwidth]{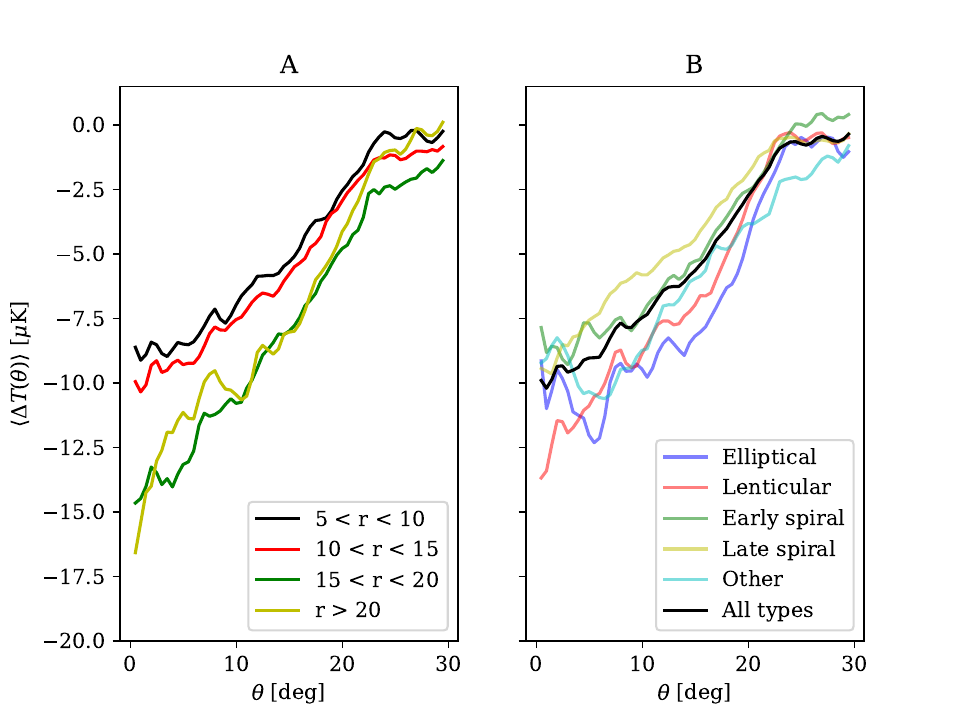}
  \caption{Panel A: Mean radial profiles for data grouped for different isophotal radii ranges in kpc. Panel B: Mean radial profiles for different galaxy types. Galaxies with radial velocities up to 5000 km/s are considered in both panels. \label{fig:radial_profiles_sizes_types}}
\end{figure}
Figure~\ref{fig:radial_profiles_sizes_types}A shows an analysis of mean radial profiles categorized by the extended isophotal radii of galaxies. We used the isophotal radii values from column 20 of the dataset (as described in~\cite{Huchra2012}), converting them to kpc. Our results appear to align with the general trend observed by~\cite{Luparello2023}, suggesting that larger galaxies seem to exhibit more pronounced CMB temperature decrements. However, it is important to note that the largest galaxies in our sample do not necessarily display the most significant temperature decrement. In addition, we observe a difference in the shape of the radial profiles of galaxies with radii greater than 15 kpc compared to those smaller than 15 kpc.  The statistical significance and robustness of this size-dependent trend therefore require further evaluation. We examine the potential significance of this size-dependent trend later in this section.

Figure~\ref{fig:radial_profiles_sizes_types}B presents the mean radial profiles grouped by different galaxy types. Notably, we observe that the CMB decrement profile is consistently present across all galaxy types, not exclusively in spirals as previously thought. This finding represents a significant departure from earlier analyses. The discrepancy likely stems from the different scaling approaches used in various studies. For instance, \cite{Luparello2023} and \cite{Addison2024} employed a logarithmic scale for $\theta$ starting at $0.01$ degrees, which emphasizes smaller angular scales.
In contrast, our analysis uses a linear scale beginning at $0.2$ degrees, providing a more uniform representation across all scales and revealing the consistency of the effect across galaxy types. Although~\cite{Hansen2023} considered $\theta \geq 0.2$ degrees, they based their template construction solely on large late-type spirals, following the conclusions of~\cite{Luparello2023}. 

We quantify the magnitude of the CMB temperature decrement by calculating the area below the mean radial profile. This area, denoted as $A(\theta_\mathrm{max})$, is computed for both observational data and simulations up to a maximum angular separation from the galaxy centers, $\theta_\mathrm{max}$. The area estimator is defined as:
\begin{equation}
A(\theta_\mathrm{max}) = \sum_{i=1}^{N-1} \frac{\langle \Delta T(\theta_i)\rangle + \langle\Delta T(\theta_{i+1})\rangle}{2} \Delta\theta,
\end{equation}
where $\langle\Delta T(\theta_i)\rangle$ is the mean temperature decrement at angular distance $\theta_i$, and $N$ is the number of bins such that $\theta_N = \theta_\mathrm{max}$. This trapezoidal approximation provides an estimate of the integrated temperature decrement. 

\begin{figure}[tb]
  \centering
  \includegraphics[width=\textwidth]{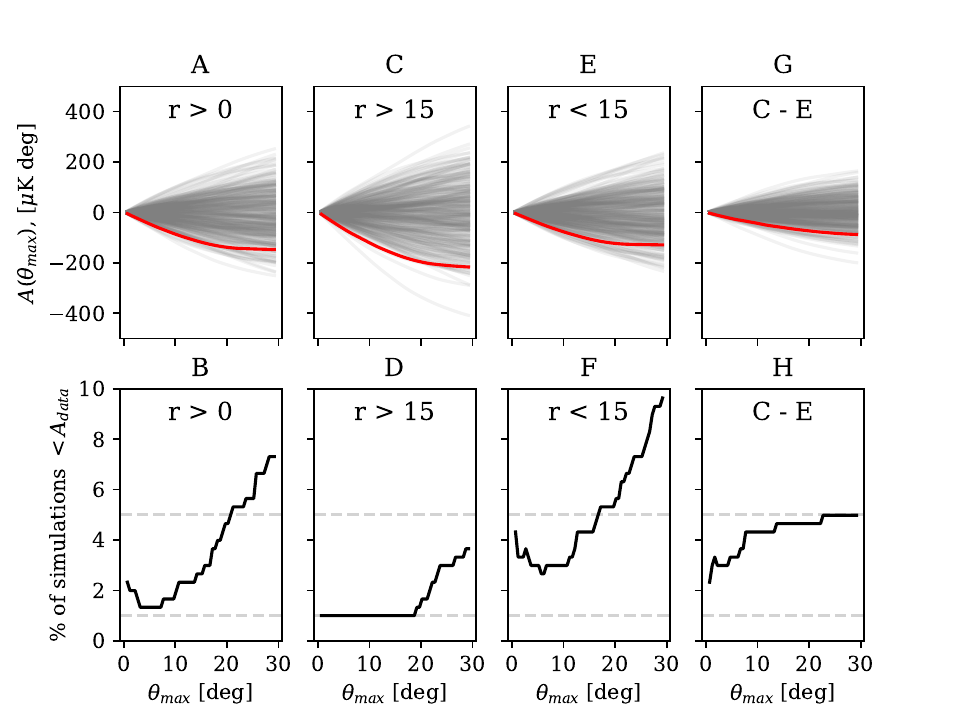}
  \caption{Panel A: Area estimator versus maximum angular distance for SMICA data (red line) and $300$ simulations (grey lines). Panel B: Percentage of SMICA simulations with lower area than the data as a function of maximum angular distance. Dashed horizontal lines are included at $1\%$ and $5\%$ as reference levels. Panels C \& D: Same as A \& B, but for galaxies with isophotal radii $r > 15$ kpc. Panels E \& F: Same as A \& B, but for galaxies with $r < 15$ kpc. Panels G \& H: Same as A \& B but for the difference of area for large galaxies ($r > 15$ kpc) minus area for small galaxies ($r < 15$ kpc).
  All panels consider only galaxies with radial velocities in the range $0 < cz < 5000$ km/s.  \label{fig:Area}}
\end{figure}
Figure~\ref{fig:Area} presents the area estimator $A(\theta_\mathrm{max})$ for the SMICA data (red line) and $300$ simulations (grey lines) across a range of maximum angular distances ($\theta_\mathrm{max}$), up to $30^\circ$. This analysis compares how the signal accumulates with increasing angular distance for several subsets: all galaxies (panel A); large galaxies with $r > 15$ kpc (panel C); small galaxies with $r < 15$ kpc (panel E); and the difference between large and small galaxies (panel G). Specifically, panel G represents the difference between the area of the average CMB radial profile around large galaxies (shown in panel C) and the area of the average CMB radial profile around small galaxies (shown in panel E). This difference allows us to directly compare the cumulative signal strength between large and small galaxies as a function of angular distance, providing insight into potential size-dependent effects in the CMB temperature decrement.

The corresponding bottom panels (B, D, F and H) show the lower tail probabilities of $A(\theta_\mathrm{max})$ for these subsets, respectively. In all four cases, we observe a significant temperature decrement in the $1\% - 5\%$ range up to $\theta_{\mathrm{max}} \approx 20^\circ$. It is worth noting that if we were to consider a two-tailed test, the $p$-values would be twice these lower tail probabilities. However, even with this more conservative approach, the results would still remain significant in the $1\% - 5\%$ range up to $\theta_{\mathrm{max}} \approx 10^\circ$. Our results indicate a more persistent and significant signal compared to the $0.2\sigma$ to $1.7\sigma$ range reported by~\cite{Addison2024}.

Regarding the size dependence analysis (panels G and H), it is important to note that while the lower tail probability falls within the $1\% - 5\%$ range, the \textit{a posteriori} choice of the $r = 15$ kpc threshold somewhat diminishes the significance of this particular result. This fact, combined with the relatively modest significance level, suggests that any conclusions about size dependence should be drawn cautiously. While there may be a hint of size-dependent effects, we refrain from making strong claims based on this analysis alone.

Regarding the galaxy type analysis presented in Figure~\ref{fig:radial_profiles_sizes_types}B, the visual impression suggests that the differences between galaxy types are not significant. To verify this observation quantitatively, we calculated the area difference between the radial profiles of elliptical and late-type spiral galaxies, which exhibit the highest and lowest areas respectively. We then compared this difference to the same calculations performed on our simulations. Our analysis confirms that the observed differences between galaxy types are not statistically significant. This finding is consistent with our earlier observation that the CMB decrement profile is present across all galaxy types, contrary to previous studies that suggested a predominant effect in spiral galaxies. 

The persistence of the signal across various galaxy types and sizes, combined with its extension over substantial angular distances, points to a possible underlying correlation between the CMB temperature and the local matter density field. In the following sections, we conduct a cross-correlation analysis to quantify and characterize this potential connection.

\section{Matter density and CMB cross-correlation}
\label{sec:crosscorrelation}

\subsection{Projected density maps and radial profiles}
Building upon our findings from the galaxy-based analysis, we now explore the possible cross-correlation between the CMB temperature and the local matter density field.
To quantify this relationship, we utilize the CORAS \cite{Lilow2021} reconstructed normalized density contrast, as detailed in Section \ref{sec:data}. This data is used to generate a series of HEALPix maps representing the projected matter density across four distance bins: $0-50$ Mpc/$h$, $50-100$ Mpc/$h$, $100-200$ Mpc/$h$, and $0-200$ Mpc/$h$. This approach allows us to examine the dependence of the CMB-matter density correlation  on the considered distance scale. 

To accurately represent the angular resolution of our data, we consider the spatial scale of our voxels in relation to their angular size at different distances.
At the maximum distance of 200 Mpc/h, a 2 Mpc/h voxel subtends an angle of approximately 0.6 degrees. Hence, we employ a HEALPix resolution parameter of $N_{\mathrm{side}} = 256$ for our projected density maps, resulting in a mean pixel spacing of about 0.2 degrees. To construct these maps, we add the density contrast for all voxels intersecting each line of sight (represented by unit vectors corresponding to CMB map pixels) up to the specified distance cut-off. 
To ensure consistency with the CMB temperature anisotropy maps, which are already corrected for monopole and dipole contributions, we remove these components from our projected density maps as well. 

\begin{figure}[tb]
  \centering
  \includegraphics[width=\textwidth]{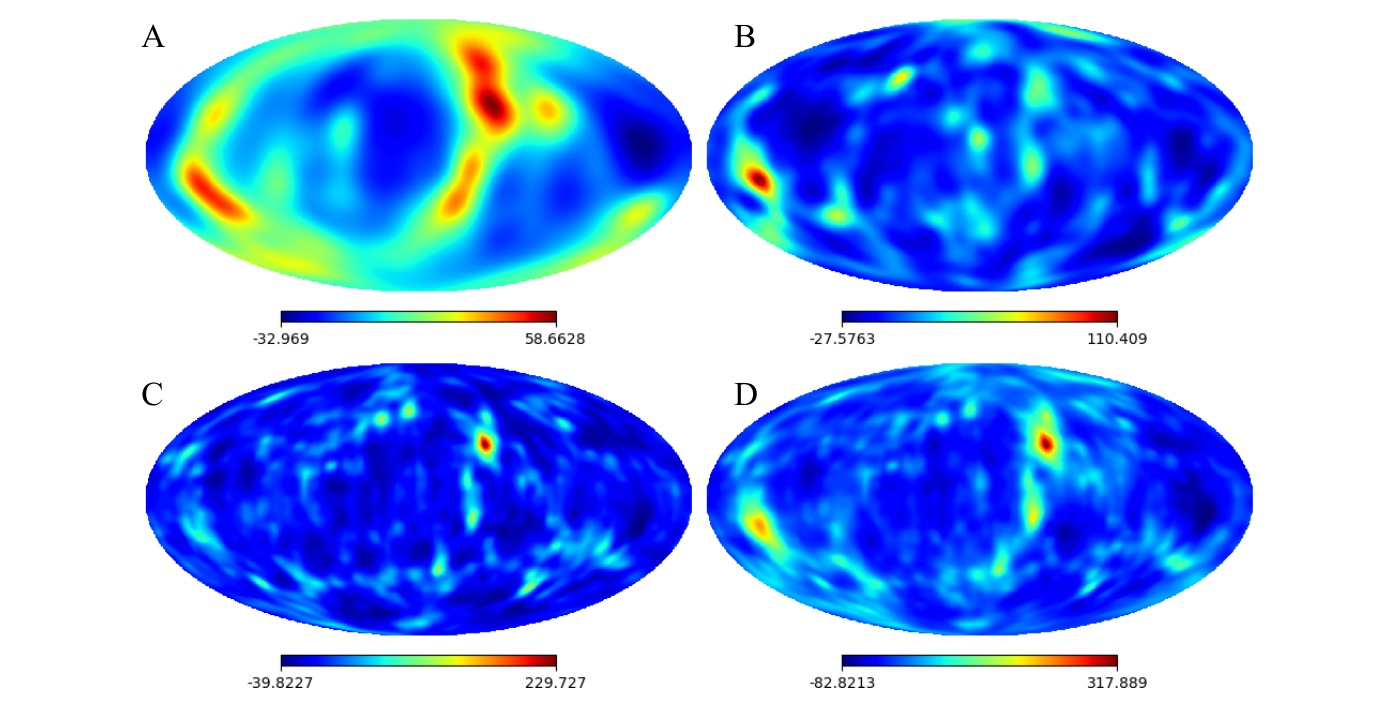}
  \caption{
  Projected matter density maps derived from the 2MASS Redshift Survey for different distance bins: (A) $0-50$ Mpc/$h$, (B) $50-100$ Mpc/$h$, (C) $100-200$ Mpc/$h$, and (D) $0-200$ Mpc/$h$. Monopole and dipole have been removed from the maps. 
  \label{fig:projected_density}}
\end{figure}

Figure~\ref{fig:projected_density} shows the projected density maps smoothed with a Gaussian beam to mitigate grid effects introduced by the voxelization of the density field. 
For the $0-50$ Mpc/$h$ bin, which approximately corresponds to the radial velocity range $0 < cz < 5000$ km/s,  we employed a Gaussian beam with a Full Width at Half Maximum (FWHM) of $9.16$ degrees. This beam size corresponds to the angular extent of about $4$ voxels at $25$ Mpc/$h$, making it suitable for our analysis as it matches the characteristic angular scale of structures in the projected density map. 
While smaller beam sizes might introduce spurious correlations due to these grid effects, we verified that our results remain consistent across various smoothing scales. For instance, using a Gaussian beam of $4.58$ degrees FWHM yielded similar outcomes.
We extended this approach to the other distance bins, verifying result consistency across Gaussian beams with FWHM ranging from $2^\circ$ to $5^\circ$. 

We now proceed to explore the correlation between the CMB temperature and the local matter density distribution using the projected density maps. 
To analyze this relationship, we adapt our earlier methodology, calculating mean radial profiles using Eq.~\eqref{eq:RadialProfile}. However, instead of centering on galaxy positions, we now focus on local maxima and minima in the projected density map. 

\begin{figure}[tb]
  \centering
  \includegraphics[width=\textwidth]{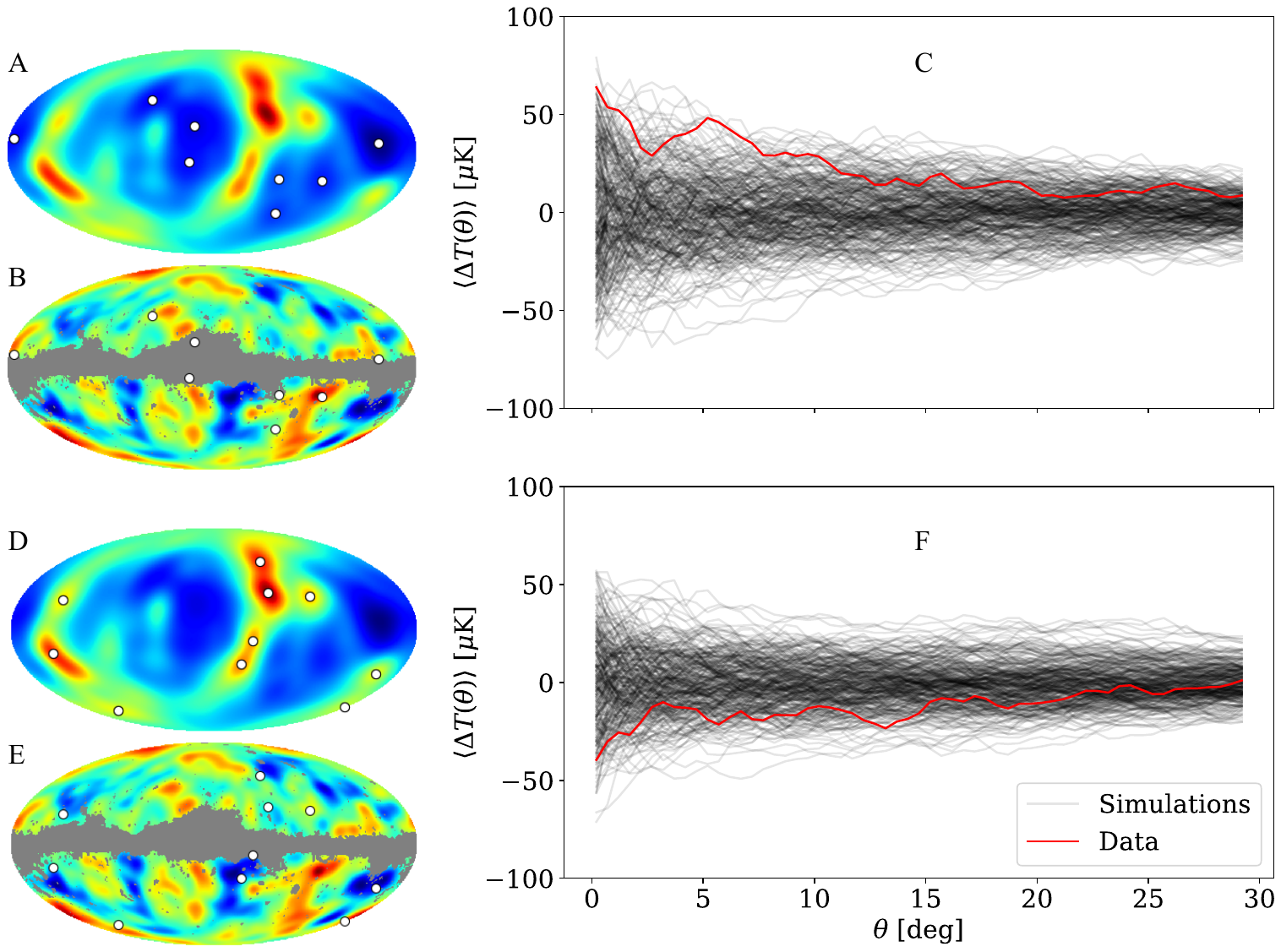}
  \caption{Analysis of CMB temperature at projected matter density extrema. Panel A: Projected matter density map for distances up to $50$ Mpc/$h$ (as in Figure~\ref{fig:projected_density}A) with local minima below $-\sigma$ marked as white dots. Panel B: SMICA CMB map smoothed with FWHM = 9.16 degrees, with common mask (grey) and local minima positions from A (white dots). Panel C: Mean radial CMB temperature profile at local minima positions. Red line shows SMICA data; grey lines represent 300 SMICA simulations. Panels D-F: Analogous to Panels A-C, but for local maxima above $+\sigma$ in the projected matter density map. 
  \label{fig:density_profiles}}
\end{figure}
Figures~\ref{fig:density_profiles}A and B illustrate the local minima at threshold $-\sigma$, while Figures~\ref{fig:density_profiles}D and E depict the local maxima at threshold $+\sigma$, where $\sigma$ is the standard deviation of the projected density map outside the common mask.
If a negative correlation exists between the CMB and matter density, we would anticipate that local density maxima would yield a negative mean radial profile in the CMB. In contrast, local density minima would be expected to produce a positive mean radial profile in the CMB. 

As demonstrated in Figures~\ref{fig:density_profiles}C and F, our data indeed exhibits this pattern. The mean radial profile is negative for density maxima and positive for density minima, suggesting a potential anticorrelation between matter density and the CMB. Note that only the results for the $0-50$ Mpc/$h$ density map are shown since we did not find this pattern in any other considered distance range, in concordance with the galaxy-based analysis.

\subsection{Cross-correlation analysis}
To further investigate this relationship and quantify its significance, we proceed with a cross-correlation analysis between the CMB temperature and the projected matter density field in harmonic space. This approach allows us to quantify the relationship between these two fields across different angular scales. The cross-correlation is defined as:
\begin{equation}
  X_{\ell} = \frac{1}{2\ell + 1} \sum_{m=-\ell}^{\ell} a_{\ell m}^\mathrm{CMB} a_{\ell m}^{\mathrm{density}*},
\end{equation}
where $a_{\ell m}^{\mathrm{CMB}}$ and $a_{\ell m}^{\mathrm{density}}$ represent the spherical harmonic coefficients of the CMB temperature map and the projected density field, respectively. The asterisk denotes complex conjugation.

We perform our analysis using a HEALPix resolution of $N_{\mathrm{side}} = 64$ and apply the common mask to both the CMB and density maps. To account for mask-induced correlations between multipoles, we employ the pseudo-$C_{\ell}$ estimator provided by NaMaster~\cite{Alonso2019} for computing the cross-correlation.

To quantify the significance of the cross-correlation, we define the cumulative sum of the normalized cross-correlation statistic:
\begin{equation}
S(\ell_{\mathrm{max}}) = \sum_{\ell=2}^{\ell_{\mathrm{max}}} \frac{X_{\ell}}{\sigma_{\ell}},
\end{equation}
where $\sigma_{\ell}$ is the standard deviation of the cross-correlation at multipole $\ell$, calculated from the simulations. We choose this estimator for its simplicity and its ability to detect persistent negative correlations across multipoles, which is the pattern we expect in the data if a true correlation exists. In contrast, for random simulations, we anticipate that the sign of the correlation will oscillate for different multipoles, leading to a lower cumulative sum. 

\begin{figure}[tb]
  \centering
  \includegraphics[width=0.8\textwidth]{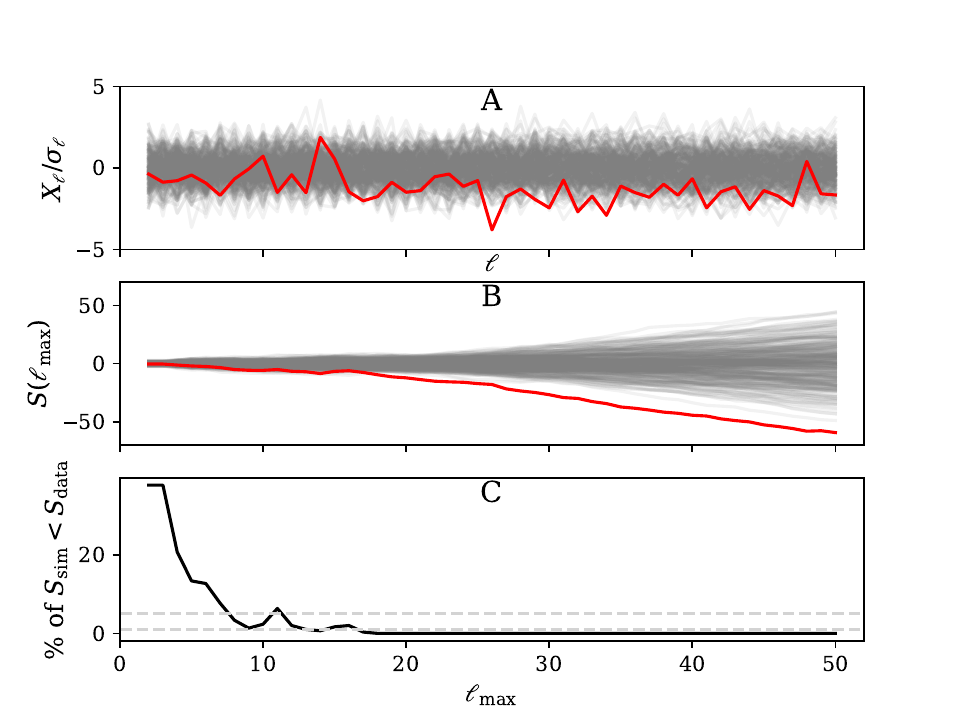}
  \caption{Cross-correlation analysis results. Panel A: Normalised CMB-projected matter density cross-correlation statistic as a function of multipole $\ell$. The red line shows SMICA data, and grey lines represent SMICA simulations. Panel B: Cumulative sum of normalised cross-correlation versus maximum multipole $\ell_{\mathrm{max}}$. Color coding as in panel A. Panel C: Lower tail probabilities for the $S(\ell_{\mathrm{max}})$ statistic. Light grey dashed lines at $1\%$ and $5\%$ are included as reference levels.
    \label{fig:crosscorrelation}}
\end{figure}
Figure~\ref{fig:crosscorrelation} shows our findings. Panel A displays the normalized cross-correlation, $X_\ell/\sigma_\ell$, for SMICA data (red line) and $300$ SMICA simulations (grey lines). Panel B shows the cumulative sum $S(\ell_{\mathrm{max}})$. Panel C presents the lower tail probabilities for $S(\ell_{\mathrm{max}})$, estimated from the simulations.

For $\ell_{\mathrm{max}} \gtrsim 15$, these probabilities fall below $1\%$, and at $\ell_{\mathrm{max}} \gtrsim 20$, no simulations yield a lower cumulative sum than the observed data. Even when considering a two-tailed test, which would double the $p$-values, these results imply a p-value below $0.7\%$.

We have verified that this result is robust across different component separation methods (SMICA, NILC, SEVEM, and Commander). However, a similar analysis using distance bins $50-100$, $100-200$ and $0-200$ Mpc/$h$ reveals no significant cross-correlation, suggesting that this signal is confined to the local universe. 

\section{Template fit}
\label{sec:TemplateFit}

\subsection{Best fit amplitude}
The analysis above reveals a significant correlation between the CMB and the projected matter density field at distances up
to $50$ Mpc/$h$. 
Although the physical origin of this correlation remains unclear, one plausible approach is to treat the projected matter
density field as a template, under the assumption that the matter density field induces a linear decrement in the CMB temperature. 

We can then perform a template fitting analysis to determine the best-fit amplitude of the template. 
Thus, we define the likelihood as follows:
\begin{eqnarray}
  \mathcal{L} & \propto & \exp \left( -\frac{1}{2} \chi^2 \right), \\
  \label{eq:Likelihood}
  \chi^2 & = & (\mathbf{D} - A \mathbf{T})^t \mathbf{\Sigma}^{-1} (\mathbf{D} - A \mathbf{T}),
\end{eqnarray}
where $\mathbf{D}$ is the CMB map, $\mathbf{T}$ is the projected matter density field, $A$ is the amplitude of the template, 
and $\Sigma$ is the covariance matrix of the CMB map. 
Under the assumption of isotropy, the covariance matrix $\mathbf{\Sigma}$ is fully defined by the \emph{Planck} angular power spectrum 
($C_\ell$):
\begin{equation}
\Sigma_{ij} = \sum_{\ell=0}^{\ell_{\text{max}}} \frac{2\ell + 1}{4\pi} C_\ell b_\ell^2 P_\ell \left( \cos \theta_{ij} \right),
\end{equation}
where $\Sigma_{ij}$ is the covariance between pixels $i$ and $j$, and $\theta_{ij}$ is the angle between them. 
$P_\ell$ denotes the Legendre polynomials, $b_\ell$ is an effective window function associated with
the $N_{\text{side}}$ resolution and beam size, and $\ell_{\text{max}}$ is the maximum multipole considered.
An uncorrelated regularization noise of $1.7 \times 10^{-2}$ $\mu$K is added to the covariance
matrix before inverting it. Regularization noise realizations are
added to the data and simulations for consistency.

In this section we use the common mask, resolution parameter $N_{\mathrm{side}} = 64$ and beam size $9.16$ degrees FWHM for data, simulations,
template and covariance matrix. 

The detectability of the template fit is sensitive to the removal of monopole and dipole components outside the mask in the CMB maps. We define "detectability" here as the ability to distinguish between CMB maps with and without the added template, quantified through our Bayes Factor analysis (detailed later in this section). Our a priori tests with simulations reveal that when we remove monopole and dipole components outside the mask, we observe lower detectability. That is, there is reduced separation between the Bayes factors of simulations with and without the added template, i.e. the power of the test is lower.
This removal may introduce inconsistencies between the data and our covariance model, as the Planck power spectrum used in our covariance matrix calculation already assumes zero monopole and dipole.

SMICA and NILC methods provide maps with monopole and dipole removed across the entire sky.
This characteristic makes them ideal for our analysis, as no further removal is necessary, and it preserves our ability to detect potential signals. 
In contrast, Commander and SEVEM maps retain these components, necessitating their removal outside the Galactic mask for consistent analysis. However, this required processing reduces detectability for these maps, as found from simulations.
Consequently, we opt to use only SMICA and NILC maps without further monopole and dipole subtraction in our subsequent analyses, as this approach maximizes our ability to detect the potential signal. It is worth noting that, based on the results from previous sections, we do not expect significant differences due to the choice of component separation method, aside from this monopole and dipole removal issue.

\begin{figure}[tb]
  \centering
  \includegraphics[width=0.8\textwidth]{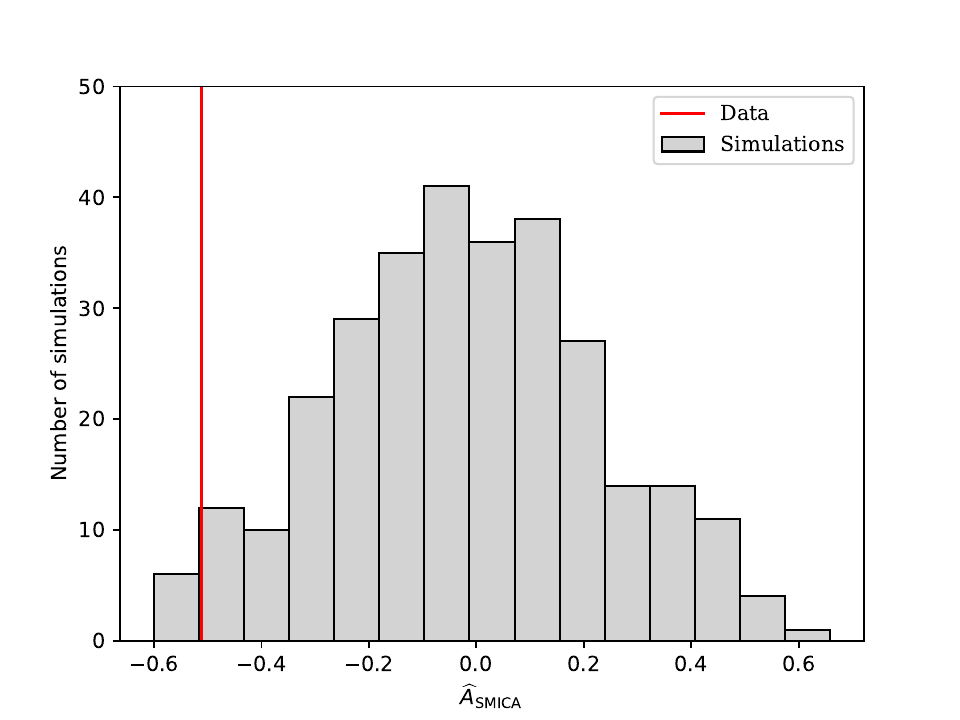}
  \caption{Best-fit amplitudes for SMICA data (red line) and simulations.
    \label{fig:BestFitA}}
\end{figure}

First, we determine the best-fit amplitude of the template for the data and simulations by minimizing $\chi^2$. This is done explicitly by differentiation, $\mathrm{d}\chi^2/\mathrm{d}A = 0$, which yields the best fit amplitude:
\begin{equation}
  \widehat{A} = \frac{\mathbf{T}^t \mathbf{\Sigma}^{-1} \mathbf{D}}{\mathbf{T}^t \mathbf{\Sigma}^{-1} \mathbf{T}}.
\end{equation}
We obtain best fit amplitudes $\widehat{A}_{\text{SMICA}} = -0.50$ and $\widehat{A}_{\text{NILC}} = -0.51$ for the SMICA and NILC maps, respectively. In Figure~\ref{fig:BestFitA}, we show the distribution of $\widehat{A}$ obtained from simulations. We find that the fraction of simulations having a lower best-fit amplitude than those of the SMICA and NILC maps are $p_{\text{SMICA}} = 2\%$ and $p_{\text{NILC}} = 1.3\%$, respectively. 
Considering the possibility of both positive and negative correlations, the two-tailed $p$-values are $p_{\text{SMICA,two-tailed}} = 4\%$ and $p_{\text{NILC,two-tailed}} = 2.6\%$. These two-tailed $p$-values account for the fact that we would consider either direction of correlation equally surprising.

In addition, the error on the amplitude can be estimated as: 
\begin{equation}
  \sigma_A = \sqrt{\left( \frac{\mathrm{d}^2 \chi^2}{\mathrm{d} A^2} \right)^{-1}} = \sqrt{\left( 2 \mathbf{T}^t \mathbf{\Sigma}^{-1} \mathbf{T} \right)^{-1}} = 0.17\,.
\end{equation}
This suggests a non-zero amplitude for the template at the level of about $3\sigma$.

\subsection{Bayes factor}
%
Here we compute the Bayes factor for both the data and the simulations. 
While we do not have a specific model, prior, or prior probability ratio 
for the models, we can proceed by defining $H_0$ as the null hypothesis (i.e., the template is not present, $A=0$) 
and $H_1$ as the alternative hypothesis (i.e., the template is present, $A \neq 0$). 
The prior probability of parameter $A$ under hypothesis $H_0$ is therefore a Dirac delta function at $A=0$.
For $H_1$ we assume a flat prior for $A$ in the range $-2$ to $2$ since the likelihood is negligible outside this range.
As the prior is not given by a specific model, the absolute value of the Bayes factor is not meaningful.
Hence, we focus on the comparison of the Bayes factor between the data and simulations (see e.g.~\cite{Amendola:2024prl} for further discussion of such mixed Frequentist-Bayesian approaches). 

The Bayesian evidence for the template is defined as:
\begin{equation}
  \mathcal{Z} = \int \mathcal{L}(A) \pi(A) \, \mathrm{d}A,
\end{equation}
where $\pi(A)$ is the prior probability density for the amplitude, and  $\mathcal{L}$ is the likelihood defined in Eq.~\eqref{eq:Likelihood}.
In our case the integral is discretized and the evidence is calculated as the sum of the likelihood 
times the prior for each amplitude value:
\begin{equation}
  \mathcal{Z} = \sum_i \mathcal{L}(A_i) \pi(A_i) \Delta A,
\end{equation}
where the sum is performed over the considered amplitude values $A_i$. 
The Bayes factor, which is the ratio of the evidences for the two hypotheses, is given by:
\begin{equation}
  B = \frac{\mathcal{Z}_1}{\mathcal{Z}_0},
\end{equation}
where $\mathcal{Z}_1$ is the evidence for $H_1$ and $\mathcal{Z}_0$ is the evidence for $H_0$.

\begin{figure}[tb]
\centering
\includegraphics[width=0.8\textwidth]{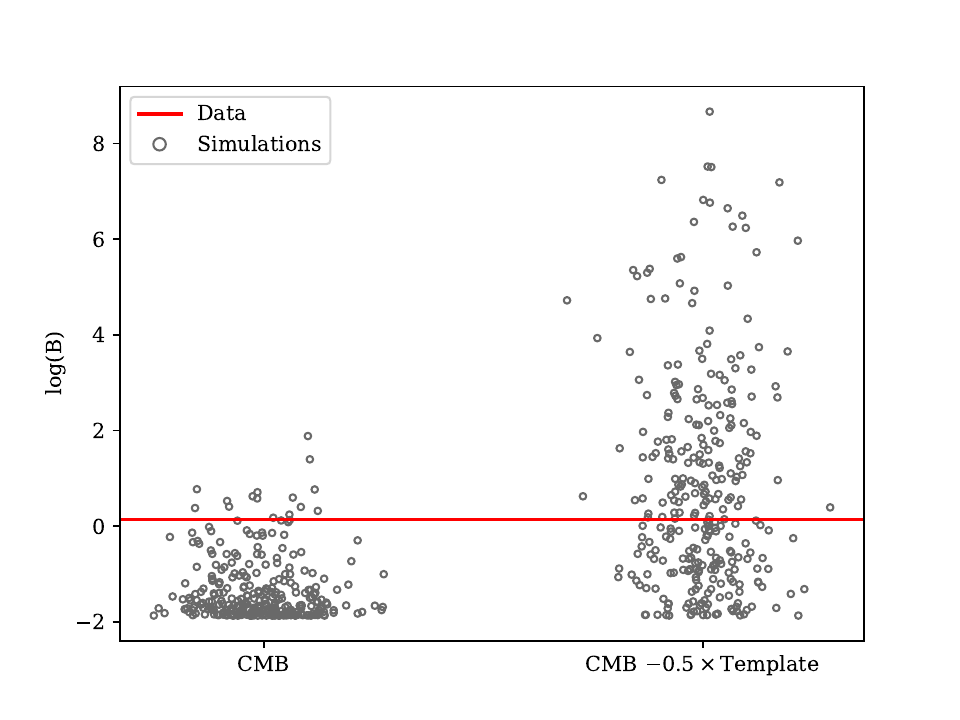}
\caption{The circles represent the log Bayes factor for 300 SMICA PR3 simulations without template added (left) and with template times best-fit amplitude ($\widehat{A}_{\text{SMICA}} = -0.50$) added (right). 
  The data is shown as a red horizontal line. Small random noise has been added to the x-coordinates to improve the visibility of overlapping data points. \label{fig:Bayes_SMICA}}
\end{figure}

We calculate the Bayes factor for the data and simulations using the SMICA (Figure~\ref{fig:Bayes_SMICA}) and NILC maps, 
finding $p_{\text{SMICA}} = 5\%$ and $p_{\text{NILC}} = 4\%$ of simulations having a higher Bayes factor than the data. Moreover, to test the detectability of the template fit, we add a template to the simulations with an amplitude of $-0.5$ and
perform the same analysis. We find that the template has $p_{\text{SMICA}} = 54.3\%$ and $p_{\text{NILC}} = 49\%$ of simulations
with higher Bayes factor than the data for this set of simulations.  
This indicates a better consistency of the data with the alternative hypothesis, i.e.~with the template added to the simulations, than without the template.

\section{Discussion}
\label{sec:Discussion}

In the previous sections, we have quantified the significance of the CMB decrement in the directions of nearby galaxies, with $p$-values in the $1-5\%$ range. We have also examined this decrement as a correlation between the CMB and the local matter density, finding a significance at the level of $<0.7\%$. While the possibility that this is a statistical fluctuation cannot be discarded, it is useful to explore possible physical origins for this effect. In order to guide this discussion, we highlight several key properties of the CMB decrement which are apparent from the analysis above:
\begin{enumerate}[label=(\roman*)]
\item The decrement is independent of the component separation method used;
\item the effect is present for all galaxy types, and is correlated with the projected matter density field;
\item the effect extends over large angular distances from the centres of galaxies;
\item the effect is approximately independent of the frequency of the CMB map.
\end{enumerate}

These results provide useful hints about the origin of the effect. Point (i) suggests that the effect is not due to galactic foregrounds, otherwise the choice of component separation method would impact the results. Point (ii) suggests that the effect does not depend strongly on any specific physical properties of the galaxies (otherwise we might expect the decrement to vary between galaxies with different structures). Indeed, we have seen that the CMB decrement is correlated with the reconstructed local matter density. This motivates us to consider a purely gravitational origin for the decrement. For example, the Integrated Sachs-Wolfe (ISW) Effect~\cite{1967ApJ...147...73S,1968Natur.217..511R,Cooray:2002ee} 
 arises from the propagation of CMB photons through the evolving gravitational potential. The linear ISW effect arises due to the late time acceleration of the Universe (peaking at low multipoles), while the non-linear ISW effect is sourced by the non-linear evolution of structure. However, the size of the effect is expected to be at least an order of magnitude smaller than that observed here~\cite{Cai:2008sm,Watson2014}. Moreover, for multipoles $\ell < 50$ a positive temperature-density correlation is expected~\cite{Watson2014}, in contrast to the anti-correlation found here.

Alternatively, Ref.~\cite{Addison2024} suggested that the kinematic Sunyaev Zel’dovich effect (kSZ) effect~\cite{1980MNRAS.190..413S} may play a role in the decrement. The kSZ effect arises from the scattering of CMB photons off free electrons, giving rise to a Doppler boosting if the electrons have non-zero bulk velocity. This bulk velocity could in principle be provided by the bulk velocities of galaxies in the local Universe. However, given that we have found no obvious directional dependence of the CMB deficit, this would require a coherent, isotropic motion of the galaxies within a distance 50 Mpc/$h$, which seems implausible, requiring that the Milky Way happens to be at the centre of some bulk flow.

In light of the fact that these more standard processes cannot explain the observed CMB decrement, we now explore whether the effect could be due to interactions between CMB photons and Dark Matter (DM) in the halos of galaxies. Indeed, the fact that the effect is not limited to the centres of galaxies but extends to large angular distances (iii) suggests that instead of the baryonic density, it may be the \textit{Dark Matter}  density which is most relevant. 

If indeed the decrement is due to the interactions of CMB photons with DM, this interaction may take the form of some scattering or absorption process~\cite{Kavanagh:2018xeh,Arina:2020mxo}. For example, CMB photons may undergo Thomson-like scattering off DM in galaxies (for example, if the DM is `milli-charged', consisting of heavy particles with an electric charge much smaller than that of the electron~\cite{McDermott:2010pa,Berlin:2018sjs}). CMB photons initially traveling along the line-of-sight would be scattered away, while photons which would not otherwise reach the observer may be scattered into the line-of-sight. Because Thomson scattering is isotropic, some fraction of the photon flux in a given direction would be replaced with the sky-averaged photon flux. As such, over-densities along the line-of-sight would tend to push the CMB temperature in that direction towards the mean CMB temperature $\bar{T}_\mathrm{CMB}$ in a process known as `Patchy Screening'~\cite{Dvorkin:2008tf,ACT:2024rue}. Stacking the signals from multiple directions would therefore not lead to a net decrement in the CMB temperature, suggesting that such elastic scattering processes cannot explain the observed effect.

Another possibility to consider is that the decrement could be due to the conversion of CMB photons into axions or axion-like particles (see e.g.~Refs.~\cite{Mirizzi:2006zy,Mukherjee:2018oeb,Carenza:2021alz,Mondino:2024rif,Mehta:2024wfo}). In the presence of magnetic fields, photons can be converted into pseudo-scalar particles such as axions via the Primakoff-Raffelt effect~\cite{primakoff_1951,raffelt_1988}. We could therefore envisage a loss of CMB photons from the line-of-sight through conversion to axions in the large-scale, coherent magnetic fields of nearby galaxies. However, this effect is strongly frequency dependent (with the conversion probability scaling roughly as $f^{7/2}$), in tension with the roughly frequency-independent effect we describe here (iv). 

We note also that the effect is only observed for galaxies below 50 Mpc/$h$. This may be because the uncertainties on the density at greater distances are too large to confidently observe the effect. If this `cut-off' at 50 Mpc/$h$ is physical, however, it would suggest that the origin of the decrement is not simply related to the line-of-sight density. For example, the effect may be related to galaxy properties which are special to galaxies in the local Universe. However, point (ii) above would seem to contradict this, not to mention the fact that there is no drastic change in galaxy evolution around $z\sim 0.02$ which would explain the sudden appearance of the effect at low redshift. Another possibility is that the effect may also correlate with the \textit{velocity} of galaxies in the local Universe. As in the discussion of the kSZ effect above, it seems implausible that the bulk motion of local galaxies should give rise to such a coherent effect. However, future studies should investigate possible cross-correlations between the CMB and local galaxy velocities.

These arguments suggest an as-yet unidentified process which depletes CMB photons along lines of sight with large matter over-densities. While we have argued that the elastic scattering of photons cannot explain this effect, this leaves open the possibility of \textit{inelastic} scattering with DM. As a concrete example, if DM has a millicharge but its rest-mass energy $m_\mathrm{DM}c^2$ is much less than the typical energy of CMB photons $E_\mathrm{CMB} \sim 10^{-3}\,\mathrm{eV}$, then these photons can undergo Compton-like scattering off the light DM particles. In this case, the scattered CMB photons may lose enough energy that they no longer lie in the CMB spectrum, leading to a decrement. We note however that models of ultralight millicharged DM are strongly constrained by stellar cooling bounds~\cite{Essig:2013lka}. Clearly, a more thorough exploration of DM models is warranted to understand the possible origins of the correlation between the CMB and local matter density.

\section{Conclusion}
\label{sec:conclusions}
In this study, we have replicated and expanded upon the analysis of the CMB temperature decrement observed in the direction of local galaxies from the 2MASS Redshift Survey. Our findings on mean radial profiles align with previous research~\cite{Luparello2023}, revealing a $10$ $\mu$K reduction in CMB temperature around galaxies with radial velocities up to $5000$ km/s. This decrement is statistically significant at approximately the $2\%$ level and appears more pronounced for larger galaxies.
The temperature decrement extends over several degrees on the sky and, contrary to earlier reports focusing solely on spiral galaxies, is evident across all galaxy types in our sample.

This observation led us to hypothesize a negative correlation between the CMB temperature decrement and the matter density field. 
To investigate this potential relationship, we generated a projected matter density field using data from the 2MASS Redshift Survey. Specifically, we utilized a constrained realization of the density field provided by~\cite{Lilow2021}. We then calculated CMB radial profiles by averaging over the positions of local extrema in this projected matter density field. The resulting profiles further corroborated our initial hypothesis.

To rigorously assess the significance of the correlation between the CMB temperature decrement and the projected matter density field, we have conducted three distinct analytical approaches. 
These methods were designed to minimize the number of a posteriori selected parameters, enhancing the robustness of our findings. By focusing on matter density, we eliminate the need for parameters such as galaxy size, type, or environment (high or low density), which were employed in previous studies.

First, we performed a cross-correlation analysis in harmonic space. Second, we treated the projected matter density field as a template and conducted a template fitting analysis. Third, we calculated the Bayes factor for the hypothesis that the template is present. 

These complementary analyses offer a multi-faceted evaluation of the proposed correlation, allowing us to assess its statistical significance from different perspectives.
Our three analytical approaches yield consistent results, indicating a correlation between the CMB temperature decrement and the projected matter density field at distances below 50 Mpc/$h$.  The cross-correlation analysis reveals a negative correlation at significance level $< 0.7\%$. The template fitting and Bayes factor calculations, while supporting this finding, yield more conservative significance levels in the $1\% - 5\%$ range. 

Consistent with previous studies, our analysis reveals that the signal loses significance for distances beyond 50 Mpc/$h$. Ref.~\cite{Hansen2023} attributes this to the diminishing angular size of galaxies at greater distances, suggesting they contribute minimally to the detected signal. However, this explanation does not fully account for matter density fluctuations, which we observed to maintain structure at angular scales of several degrees even at larger distances. A more plausible explanation lies in the properties of the reconstructed normalized density contrast. As demonstrated in Figure 9 of~\cite{Lilow2021}, the standard deviation of the density contrast increases substantially beyond 50 Mpc/$h$. This growing uncertainty in density estimation at larger distances could explain the observed lack of correlation between the CMB and matter density in these regions. 

The physical origin of the observed CMB decrement remains an intriguing open question. Our analysis has shown that the decrement is independent of both the frequency of the CMB map and the component separation method employed. This robustness across frequencies and methodologies strongly suggests that Galactic foregrounds are not responsible for the observed signal.

We have systematically ruled out several potential mechanisms such as the ISW and kSZ effects.
The large angular scale over which the correlation is found, points to a potential connection with local large-scale structure. Specifically, the dark matter distribution in galactic halos emerges as a plausible source for this phenomenon. However, we have discussed that Thomson-like scattering with dark  matter particles and conversion of CMB photons into axions are unable to explain the observed effect. Other possibilities include Compton-like scattering caused by ultralight, millicharged dark matter, but further exploration of the landscape and parameter space of possible models is required.

While the statistical significance of our findings  is not conclusive, the substantial cosmological implications of this potential CMB-matter density correlation cannot be overstated, warranting further investigation to confirm or refute this intriguing phenomenon.

\acknowledgments

Some of the results in this paper have been derived using healpy \cite{Healpy2019}, HEALPix \cite{Gorski2005}, Python3 \cite{python2009}, matplotlib \cite{Matplotlib2007}, numpy \cite{numpy2011}, astropy \cite{Astropy2018}.
This work was partially supported by the projects of the Plan I+D+I (area
of Space Research), Agencia Estatal de Investigación (AEI) with references PID2022-139223OB-C21 and PID2022-138896NB-C53. We acknowledge R.B. Barreiro for providing the 44GHz SEVEM DX9 clean map, J.L. Bernal for suggesting the use of the CORAS data and P. Vielva, F.K. Hansen, and D. García Lambas for useful discussions. The authors also thank the ‘Dark Collaboration at IFCA’
working group for useful discussions.




\providecommand{\href}[2]{#2}\begingroup\raggedright\endgroup







\end{document}